\documentclass[epj]{svjour}
\usepackage{graphics}
\newcommand{\be}{\begin{equation}}
\newcommand{\ee}{\end{equation}}
\begin{document}
\title{High temperature superconductivity from the
two-dimensional semiconductors without magnetism}
\author{Mahito Kohmoto\inst{1}, Iksoo Chang\inst{2},\and 
Jacques Friedel\inst{3}
}                     
%
%
\institute{Institute for Solid State Physics,
University of Tokyo, 7-22-1 Roppongi, Minato-ku, Tokyo, Japan
\and Department of Physics, Pusan National University, Pusan
609-735, Korea \and Laboratoir\`{e} de Physique des Solides, 
Universit\'{e} Paris-Sud, Centre d'Orsay, 91405 Orsay Cedex, France
(unit associated to the CNRS)} 
\date{Received: date / Revised version: date}
%
\abstract{
We examine the possibility of high temperature
superconductivity from two-dimensional semiconductor without antiferromagnetic
fluctuations. The weak coupling BCS theory is applied, especially where
the Fermi level is near the bottom of the conduction band. Due to screening,
the
attractive interaction is local in $k$-space. The density of states(DOS)
does not have
a peak near the bottom of the band, but
$k$-dependent contribution to DOS has a  diverging peak at the bottom. These
features lead to high temperature superconductivity which may explain the
possible
superconductivity of WO$_3$.
} 
%
\authorrunning{M. Kohmoto, I. Chang, and J. Friedel}
\titlerunning{High $T_c$ superconductivity without magnetism} 
\maketitle
One of the most spectacular discoveries in condensed matter physics is
undoubtedly the
high $T_c$ cuprates superconductors(HTSC)\cite{BM}. Many properties are not
simply
described by the standard classical BCS theory\cite{BCS}. The common
features of the
HTSC are:

(a) high superconductive transition $T_c \sim 100$K.

(b) quasi two-dimensionality, with weakly coupled CuO$_2$ layers

(c) antiferromagnetic(AF) fluctuations. The mother materials of HTSC are
insulators
with AF order below $T_N$.

(d) anisotropic superconductive gaps.

In order to understand the mechanism of HTSC,
there has been efforts to find compounds in which  Cu is replaced by an
other atom.

The compound Ru$_2$SrO$_4$ has the same lattice
structure as the HTSC (La, BA)$_2$CuO$_4$, but Cu is replaced by Ru. It is
found to
have a superconductive phase, however $T_c \simeq 1.5$K is rather low. There
are
some experimental results (for example,
NMR\cite{nmr} and muon spin resonance\cite{muon} ) to support triplet
pairings. The band structure is considerably different from that of typical
HTSC. Thus
superconductive nature of this layered compound does not seem to be  related
to HTSC.

Recently the signs of high superconductive transition temperature are
reported for
some doped semiconductors. The Na$_x$WO$_{3-x}$ sodium tungsten bronzes,
formed by
doping the insulating host WO$_3$ with Na, are $n$-type semiconductor for
$x<0.3$.
The sample with $x=0.05$  shows a
sharp metal to insulator transition as temperature is lowered below 100K.
It is
followed by a sharp drop of resistivity at $91$K as temperature is further
lowered,
it shows  signs of superconductivity at 91K.  At the same
temperature this compound exhibits a sharp diamagnetic step in
magnetization\cite{rt}. This
compound is quasi
two-dimensional, but does not show any sign of antiferromagnetic
fluctuations in
sharp contrast to the cuprate high
$T_c$ superconductors.
Possible superconductive transition  is further supported by subsequent
measurements
of Electron Spin Resonance\cite{esr}.

There is no sign of  antiferromagnetic fluctuations in
this compound  in sharp contrast to cuprate HTSC, namely the feature (c)
above does
not apply. This implies that either the mechanism of superconductivity of
WO$_3$ is
different from the cuprate superconductors or the magnetism does not play an
essential
role in the high $T_c$  cuprates.

The 5$d$-transition oxides WO$_3$ and Na$_x$WO$_3$ have nearly the
perovskite crystal with W ions occupying the octahedral cation
sites. Stoichiometric WO$_3$ is an insulator since the W 5$d$ band is empty;
when Na
ions are added to WO$_3$, they donate their 3$s$ electron to the W 5$d$ band,
resulting in bulk metallic behavior for $x
\geq 0.3$. For $x<0.3$ the Na$_x$WO$_3$ sodium tungsten, is an n-type {\em
semiconductor} (electron doped). This indicates that there exist two  bands:
the
valence band and the conduction band with the gap $G$ in between.  The
effect of
doping of Na is to  add electrons in the conduction band, thus raising the
chemical potential $\mu$.

Although the superconductivity of WO$_3$ has not been confirmed solidly and it
requires further experimental supports,  the  properties of WO$_3$ motivated
us to examine the possibility of high
$T_c$ superconductivity from  semiconductors {\em without} antiferromagnetic
fluctuations.    For simplicity,  we neglect possible contributions from the
valence
band. The conduction band is modeled by

\begin{equation}
\varepsilon({\bf k}) \simeq -2 t (\cos k_x + \cos k_y ),
\end{equation}
where $t$ is the transfer integral of the tight-binding model on the
square lattice.
This band has the  van Hove singularity in the density of
states(DOS) at half-filling.  There are a number of
works\cite{hs,friedel,newns,abrikosov,bok,fk} that try to explain high
$T_c$ cuprates by van Hove singularity. Here the present situation is totally
different since we consider a nearly empty band.

The density of states is given by
\begin{equation}
N(E)=\int \frac{1}{|\bigtriangledown
\varepsilon({\bf k})|} dl = \int \frac{1}{v} dl,
\label{DOS}
\end{equation}
where $l$ is the Fermi surface (line in two dimensions) and $v$ is the
semiclassical
velocity. For the sake of convenience we shall call $1/|\bigtriangledown
\varepsilon({\bf k})|$ as kDOS and it is plotted in Fig. 1.
Note that kDOS is diverging near the $\Gamma$ point ${\bf k}=(0,0)$ as
well as
at
$(\pm\pi,0)$ and $(0,\pm\pi)$.
This singularity is not seen in DOS since it is integrated on a vanishingly
short
Fermi line.

\begin{figure}
\resizebox{0.50\textwidth}{!}{%
  \includegraphics{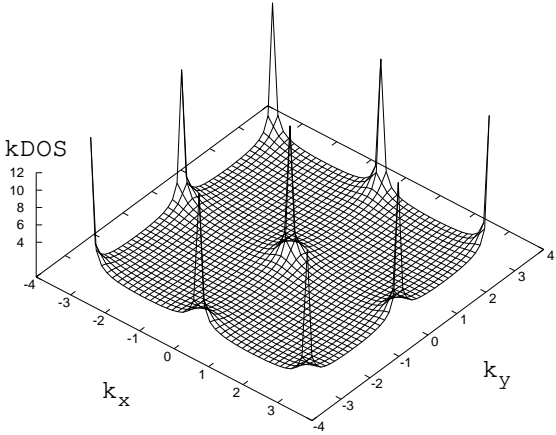}
}
\caption{\protect\footnotesize
k-dependent density of states kDOS. All the peaks are actually
diverging.
The peak at $\Gamma$ (0,0)  corresponds to  the
bottom of the band . The four peaks at ($\pm \pi, \pm \pi$) are at the top
of the
band and the others give van Hove singularities at half-filling.
}
\end{figure}

The important fact is that if interactions are local in $k$-space, kDOS
has to be considered  carefully.

With the properties of the band above in mind, we consider the gap equation,

\begin{equation}
\Delta_k = -
\sum_{k'}\frac{V_{kk'}\Delta_{k'}(T)}{2E_{k'}}\tanh\frac{E_{k'}}{2k_B T},
\label{gap}
\end{equation}
where $E_k = \sqrt{\varepsilon(k)^2 +\Delta_k (T) ^2}$, $\Delta_k(T)$ is the
gap
order parameter and $V_{kk'}$ is the interaction. The sum is restricted
within the
cutoff
$\mu -E_c<\varepsilon(k')<\mu+E_c$ where $\mu$ is the chemical potential
which is
measured from the bottom of the  band of noninteracting case.
 From the gap equation(\ref{gap}) one can obtain $T_c$ and $\Delta (T)$.
Near $T_c$,
$\Delta$ is very small, then (\ref{gap}) is linearized. $T_c$ is determined
by the
linearized equation. The gap $\Delta(T)$ is obtained by iteration of
(\ref{gap}).

If one takes the usual BCS interaction $-V$ which is constant if the two
particles are
both within the cutoff and vanishing otherwise, we have the classical BCS
result
$T_c
\sim 1.13E_c \exp (-1/NV)$, where N is the DOS near the bottom of the band.
When one
takes physically reasonable value of $V$, high $T_c$ can not be obtained
even if kDOS
is large. This is because the Fermi line is a small circle and the range of
integration is very small. It offsets the large but uniform kDOS. This
behavior is
totally analogous to the fact that the behavior of DOS --integral of kDOS
along the
Fermi line-- that is almost constant near the bottom of the band.

On the other hand if the
interaction is local in k-space, i.e. not constant in the integral, the
effect of the
large kDOS can not be canceled by the short length of the Fermi line.

In  many-body systems the interactions are mostly scr-
eened and this effect has
to be included, unless the system is exactly solvable. Thus we take a weakly
screened attractive interaction (likely to be  phonon-mediated one),

\begin{equation}
V_q = - \frac{g_q^2}{q^2 +q_0 ^2}.
\label{phonon}
\end{equation}
Here $g_q$ is the coupling constant and $q_0$ is the inverse of screening
length $L$: $q_0 \simeq  1/L$. This poor screening is supposed to be due to low
dimensions(2d). The screening length actually depends on $\mu$, but we
neglect this
effect for the sake of simplicity.
 Since the
interaction (\ref{phonon}) is local in
$k$-space, the effect of the large kDOS can not be totally canceled by the
small Fermi
line, as discussed above.

Let us first give an estimate of the effective interaction. For $\mu = -4t+
\eta$
where $\eta$ is small but larger than the cutoff $E_c$, one has, from
(\ref{gap}),

\be
2= \int_{0}^{2\pi} \int_{k'_{inf}}^{k'_{sup}}
\frac{g_q ^2 ~ \tanh \frac{k'^2t-\eta}{2k_BT_c}~~
\frac{k'}{k'^2t-\eta}}{k'^2 + k^2-2k'k\cos\theta + q_0^2}~ dk'd\theta
\ee
with $k^2t = \eta, \eta \gg E_c$ and $|k'^2t - \eta|<E_c$.
This leads to

\be
2=\pi g_q^2t \int  \frac{\tanh\kappa}{2k_BT_c}~\frac{d\kappa}{\kappa} {1
\over A}
\ee
with $k'^2t-\eta = \kappa$ and $A=\kappa^2+4\eta ~q_0^2 t+\cdots$. With the
preceding conditions, and if $\eta$ is large enough versus $E_c$, $A$
reduces to $1
/ 4\eta ~q_0^2 t $. One reverts to the BCS case of $V$ constant with

\be
V_{kk'} = -\frac{g_q^2}{q_0^2},
\ee
thus much higher $T_c$ is expected.

The numerical solution is consistent with the above analysis and gives high
critical temperature near the bottom of the band. It  is plotted in Fig.2
and it
takes the  maximum value at $\mu_{op}$ near the bottom of the band. The
decrease of
$T_c$ for
$\mu$ smaller than
$\mu_{op}$  is due to the semiconductor gap where DOS vanishes. But note
that $T_c$ does not vanish at the bottom of the band but extend to the band gap
region due to the superconductive coherence effect. It is natural that 
$T_c$
does not
depend on the cutoff in this region. The transition temperature
$T_c$ also decreases for
$\mu$ larger than  $\mu_{op}$. This is because the kDOS is getting smaller
in this
region.

\begin{figure}
\resizebox{0.50\textwidth}{!}{%
  \includegraphics{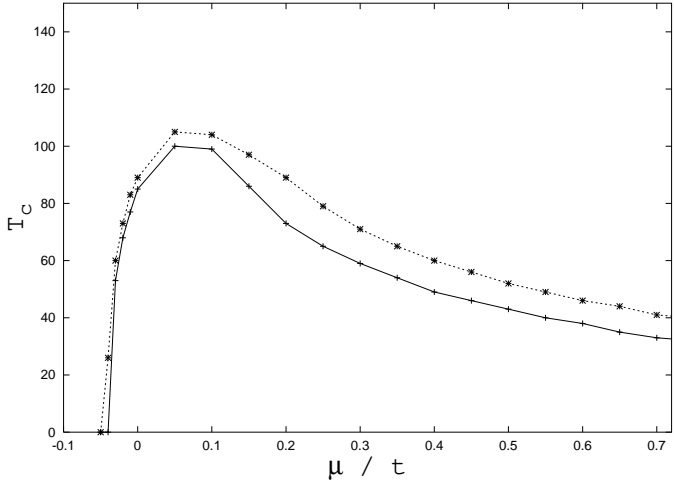}
}
\caption{\protect\footnotesize
Critial temperature as a function of $\mu/t$. The total band width is
$8t$. The
parameters are: transfer
$t=0.25$eV, screening lengs $L=15$ lattice spacings, and $g_q^2=0.6t$. The
cutoffs $E_c$
are $50$meV (top curve) and $30$meV (bottom curve).
}
\end{figure}

\begin{figure}
\resizebox{0.50\textwidth}{!}{%
  \includegraphics{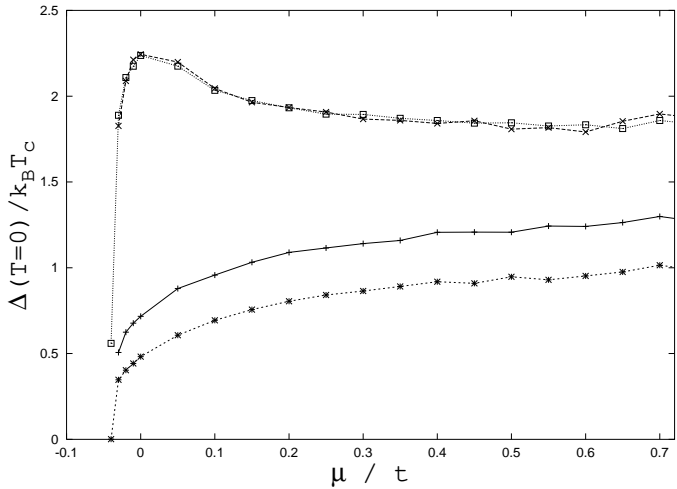}
}
\caption{\protect\footnotesize
The ratio $\Delta(0)/k_BT_c$ for the maximum and minimum of $\Delta(0)$'s. 
From top to bottom curves: the maximum ratio for $E_c = 50meV, 30meV$, and
the minimum ratio for $E_c =  30meV, 50meV$}
\end{figure}

We have a typical
$s$-wave pairings, but the gap depends on
the absolute value  $|{\bf k}|$. The maximum and minimum of
$\Delta(0) / T_c$'s are plotted in Fig.3. Note that the maxima of the gap is
almost independent of the cutoff. 
It is about $1.8$ which is close to the BCS value
$1.76$ at $\mu \simeq 0.6$. It increases as $\mu$ is lowered. On the other
hand the
minimum of the gap depends on the cut off as expected because the minimum
occurs at
the edge of the cutoff. It is less than the BCS value and decreases as $\mu$ is
decreased.

Note that
$\Delta(0)/k_B T_c$ approach zero in a similar manner to Tc. This implies that
$\Delta(0)$ decay faster than $Tc$.
If we take into account the
valence band, certainly  the behavior of
$T_c$  will   be changed. The pairings between electrons in the conduction
band and
valence band  have to be considered\cite{kt} in addition to parings between
electrons
in the same band. However this will not change the results qualitatively.

The features (c) and (d) do not apply but (a) and (b)  survive.
Therefore it can be concluded that the origin of high $T_c$ in the present
case is due to the specific features of the energy dispersion in two
dimensions.

Since we have a particle-hole duality,
these results are equally applicable to p-type semiconductors (hole doped). \\

To conclude, we study the two-dimensional conductive band with low doping.
This is motivated by some signs of a  high superconductive transition
temperature in WO$_3$:Na \cite{rt,esr}. The crucial point is the large
electron density kDOS near $\Gamma(0,0)$. We succeed in obtaining a  high
superconductive temperature ($T_c \sim 100$K) from the approximate  analysis
and the numerical computations.
This result is readily applicable to an almost full valence band.

We expect that these results are rather general. The two-dimensional
materials with  almost empty conductive band or with almost full valence
band are almost likely to have a high temperature superconductive
transition. \\

\begin {center}
{\large{\bf Acknowledgments}}
\end {center}

It is a pleasure to thank M. Takigawa for a useful discussion.
I.C is grateful to Korea-Japan binational
program through KOSEF-JSPS which made his visit to ISSP, Tokyo possible,
and to BSRI98-2412 of Ministry of
Education, Korea.




\begin{thebibliography}{}
\bibitem{ins} Here the magnitude of $G$ is irelevant, and our results will
also apply to the band insulators.
\end{thebibliography}
\end{document}